\documentclass[runningheads]{llncs}

\usepackage[T1]{fontenc}

\usepackage{graphicx}

\usepackage{rotating}
\usepackage{cite}
\usepackage{caption}
\usepackage{multirow}

\begin{document}
\title{Can LLMs Reason About Trust?}
\subtitle{A Pilot Study}
%
%
\author{Anushka Debnath\inst{1} \and
Stephen Cranefield \inst{2} \and
Emiliano Lorini\inst{3} \and
Bastin Tony Roy Savarimuthu\inst{2}}
\authorrunning{A. Debnath et al.}
%
\institute{National Institute of Technology, Durgapur, India \email{anushkadebnath77777@gmail.com}\and
University of Otago, New Zealand \email{(stephen.cranefield, tony.savarimuthu)@otago.ac.nz} \and 
IRIT, CNRS, Toulouse University, France 
\email{emiliano.lorini@irit.fr}\\
}
\maketitle              
\begin{abstract} 
In human society, trust is an essential component of social attitude that helps build and maintain long-term, healthy relationships which creates a strong foundation for cooperation, enabling individuals to work together effectively and achieve shared goals.
As many human interactions occur through electronic means such as using mobile apps, the potential arises for AI systems to assist users in understanding the social state of their relationships. In this paper we investigate the ability of Large Language Models (LLMs) 
to reason about trust between two individuals in an environment which requires fostering trust relationships. We also assess whether LLMs are capable of inducing trust by role-playing one party in a trust-based interaction and planning actions which can instil trust.

\end{abstract}
\section{Introduction}
The concept of trust is studied in  various disciplines, including sociology, psychology, business, economics and cognitive science\cite{castelfranchi2010trust, gambetta2000can, mayer1995integrative  ,ostrom2003trust, rousseau1998not, hardin2002trust}, with each domain offering different theoretical perspectives on its definition and implications \cite{castelfranchi2010trust,jdlewis}. Broadly, trust is the belief or expectation that one party (the trustor) has in another (the trustee) regarding their reliability, integrity, or competence in fulfilling a task or obligation, often based on prior experience or reputation. It emerges when the trustee’s actions align with the trustor’s goals, fostering shared purpose and mutual understanding. 
Trust promotes cooperation and reduces uncertainty in human interactions, and future advancements in computational models of trust, could play a crucial role in assisting the establishment and maintenance of relationships in computer-supported human interactions.

Within multi-agent systems (MAS), trust is typically modelled through two primary approaches: \textit{computational} models, which define trust numerically based on prior performance and reputation \cite{marshFormalisingTrustComputational1994}, and \textit{socio-cognitive} models \cite{staab2008combining}. While computational models are effective in structured environments, they fall short in capturing the complexities of goals, intentions, and context that shape trust in dynamic, real-world scenarios. In contrast, the socio-cognitive model introduced by Castelfranchi and Falcone \cite{castelfranchi2010trust,falcone2001social} asserts that trust is determined by an agent’s beliefs, goals, and desires. According to this model, trust is defined as a ``composite mental attitude'' whereby, for instance, a trustor (agent X) believes that a trustee (agent Y) possesses both the ability and the willingness to perform a specific action (A) whose execution ensures that the goal of the trustor will be achieved. However, implementing such complex mental states computationally presents significant challenges. Traditional symbolic reasoning approaches often struggle with the inherent complexity and are prone to failures in dynamic and uncontrolled environments, limiting their practical applicability \cite{lu2011neural}. We believe that LLMs show potential to fill this gap by capturing and reasoning about these nuanced mental states in a more flexible and scalable manner, as they learn trust-related patterns from large-scale data and adapt to new contexts without requiring predefined logical rules or manually crafted representations of trust, making them a promising tool for modelling trust in dynamic settings.\\
The primary objective of this research is to evaluate whether LLMs can analyse conversations between two individuals to reason about trust, thereby facilitating the establishment and maintenance of trustworthy collaborative relationships. Additionally, our study explores whether an LLM, when instructed to act as one of the individuals in a trust relationship, can generate a strategic plan of actions to foster and build trust effectively.

\section{Large Language Models} \vspace{-0.1em}Large Language Models (LLMs) are advanced AI systems designed to understand, generate, and process human language by leveraging massive datasets and transformer architectures. With billions of learned parameters, they excel in tasks such as language understanding, text generation, and general question-answering across various domains. LLMs are pretrained on vast textual data and can be fine-tuned for specific applications, enabling them to serve as versatile tools in industries like healthcare, education, finance, and software development. Their strengths include adaptability, broad coverage of diverse topics, and ease of integration into applications through APIs. Park et al. \cite{park2023generative} observed and suggested that LLMs are capable of encoding a wide range of human behaviour from their training data. In a survey of LLM agents, Xi et al. \cite{xi2023rise} explore research on how these agents engage within a societal framework. Studies in this field have examined interactions shaped by individual personalities and emotions, collaborative teamwork dynamics, and the emergence of spontaneous social behaviours.
Also, prior research \cite{savarimuthu2024harnessing, he2024norm} investigating the social reasoning abilities of LLMs has shown that they can infer when violations of social norms have occurred.

\section{The Concept of Trust}
The concept of trust has been studied in various ways. Out of the many numerous theories of trust, we chose Castelfranchi and Falcone's theory \cite{castelfranchi2010trust, falcone2001social} as the foundation for our study because it provides a comprehensive framework that captures the complexity of trust, incorporating both emotional and cognitive aspects. The theory views trust as a mental attitude that is based on beliefs and goals and, more specifically, as a positive expectation about the action of the trustee. It emphasizes on the dynamic and goal-oriented nature of trust which highlights the evolving nature of trust according to changes in situations and its link to achieving specific objectives. Trust plays a critical role in decision making because it is highly context-dependent, shaped by environmental and situational factors and guided by factors such as safety, willingness and competence. These factors make Castelfranchi and Falcone’s theory a rigorous foundation for exploring the different dimensions of trust in diverse scenarios.
According to the theory, trust has four key components as follows:
\begin{itemize}
\item 
The Trustor: The one who trusts.
\item
The Trustee: The one being trusted.
\item
The Action: The task or behaviour that the trustee is expected to perform.
\item 
The Goal: The outcome or objective that the trustor wants to achieve.
\end{itemize}
In this theory, the trustor aims to achieve a specific goal and places trust in the trustee based on the belief that the trustee possesses both the competence and the willingness to perform actions necessary for goal attainment. This conceptualization of trust
aligns with Castelfranchi and Falcone’s delegation-based model, where trust is situated within a framework of reliance on the trustee’s ability and intention to fulfill tasks that contribute directly to the trustor’s objectives.

\begin{figure}[t]    
    \includegraphics[width=\textwidth]{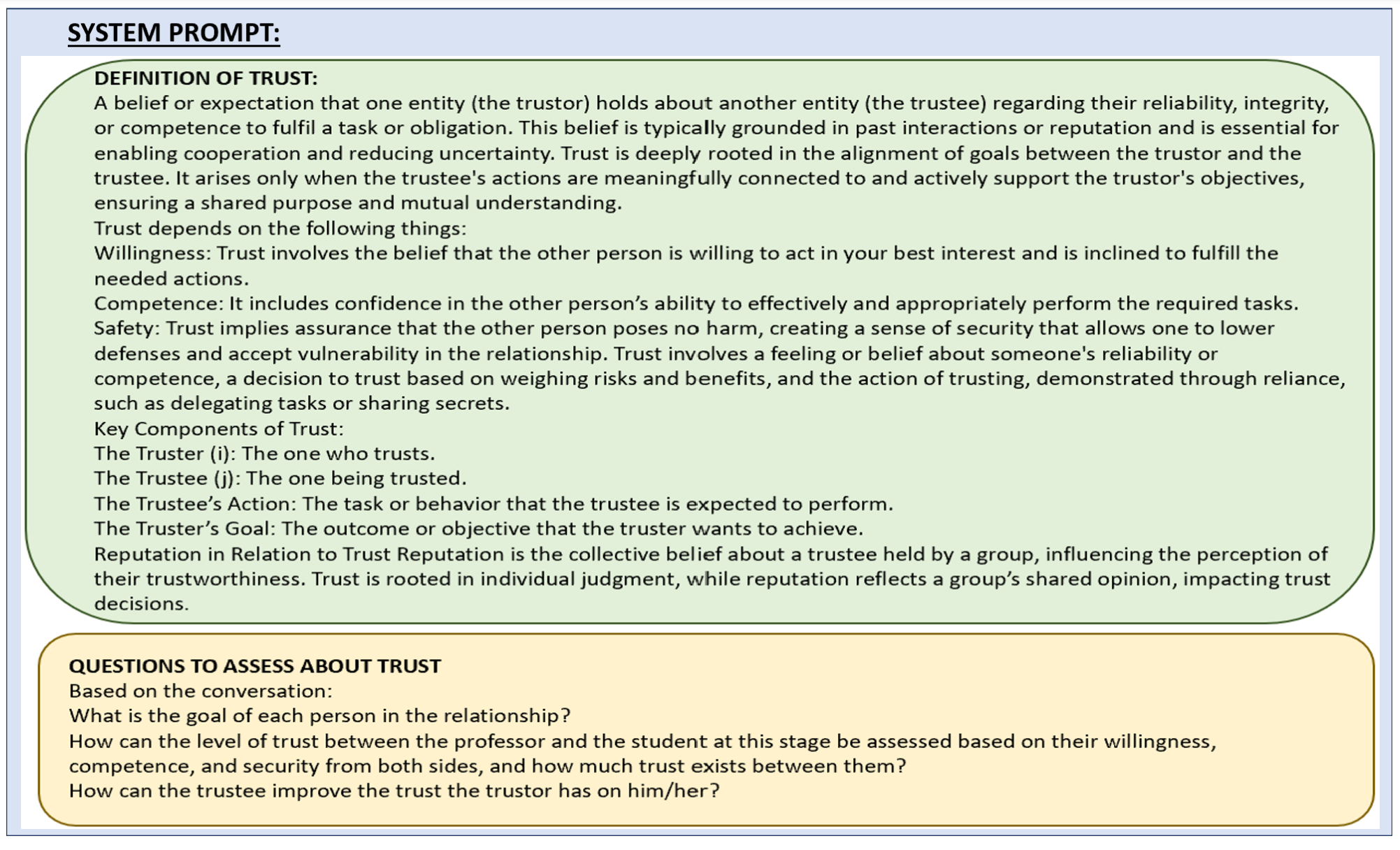}
    \caption{{System Prompt}}
    \label{system_prompt}
\end{figure}

Building on Castelfranchi and Falcone's theory, our proposal evaluates trust based on the following key factors:
\begin{itemize}
\item
Willingness: Trust involves the belief that the other person is willing to act in your best interest and is inclined to fulfill the needed actions.
\item
Competence: It includes confidence in the other person’s ability to effectively and appropriately perform the required tasks.
\item
Safety: Trust implies assurance that the other person poses no harm, creating a sense of security that allows one to lower defenses and accept vulnerability in the relationship.
\end{itemize}
In our proposal, we have identified and extracted key aspects of trust and its  meanings  from Castelfranchi and Falcone’s trust theory \cite{castelfranchi2010trust, falcone2001social}. To evaluate the capability of LLM to reason about trust, we formulated specific questions about trust to evaluate whether LLMs can effectively reason about trust based on conversational interactions between two individuals. These elements were then combined together, as illustrated in Fig.\@ \ref{system_prompt}, to construct a system prompt provided to the LLMs for  trust analysis. \par
We used four state-of-the-art commercial LLMs, namely gpt-4o, llama-3.3-70b-versatile, mixtral-8x7b-32768  and gemma2-9b-it, to assess their capability to reason about trust between two individuals in different interactions. We used 
the OpenAI API to access gpt-4o and 
Groq tools \footnote{https://groq.com/}
to access the other three LLMs. We specifically selected these LLMs to conduct our study because they were the best among their LLM families. We set the parameters to be:
\textit{top p} = $0.95$, temperature = $0.8$, and context length = $2048$\footnote{All the case studies and the LLM responses are available in the supplementary material \cite{Debnath2025}.}.

\begin{figure}[t]
    \centering\includegraphics[scale=0.4]{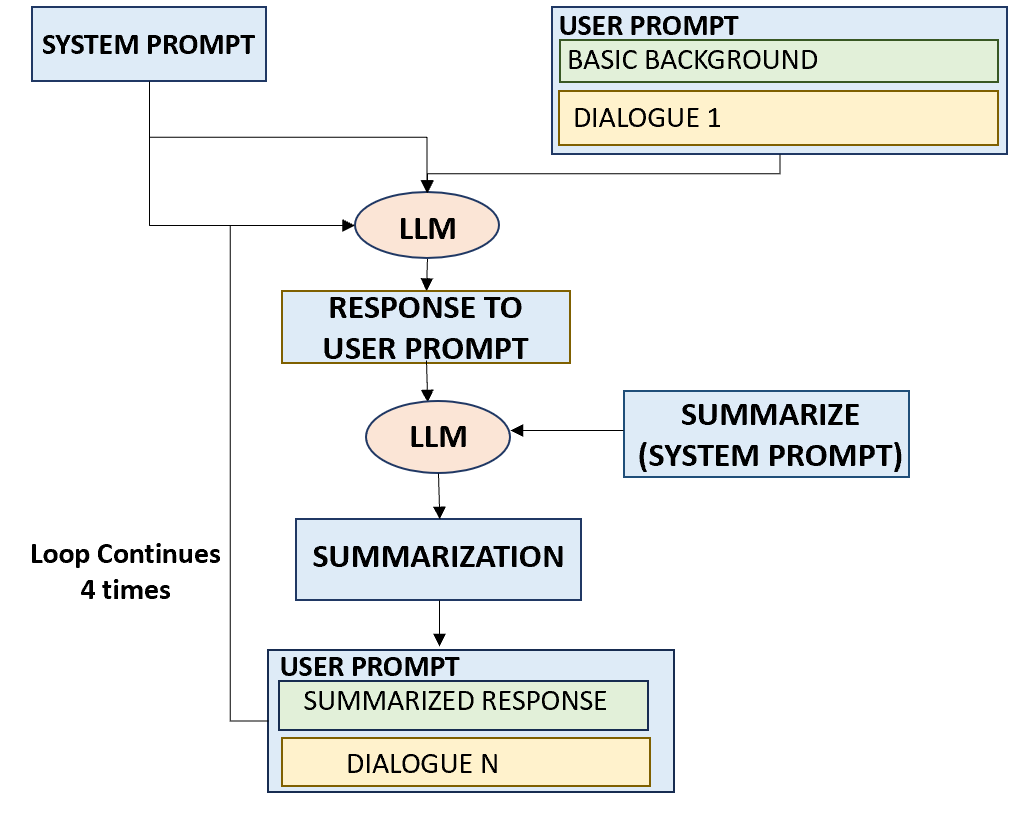}
    \caption{{Flowchart of Prompt-Response Generation}}
    \label{Flowchart}
    
\end{figure}

\section{An Application of Testing Trust Reasoning: GPT-4o Response Analysis}
In this work, we consider the relation between a PhD student and their supervisor, to assess the capability of LLMs to reason about trust. The dynamic between a PhD student and a supervisor requires a significant level of mutual trust, which must be nurtured and preserved over a long period of time to ensure successful collaboration and research outcomes. Hence, our objective is to explore whether LLMs can effectively comprehend, analyse and provide insight into the intricate and nuanced aspects of trust inherent in such long-term professional relationships.\par
In our work, we explore trust in professor-student interactions through two case studies. In the first case study, we analyse five different interactions where an LLM evaluates trust dynamics based on a sequence of exchanges. In the second case study, we examine two scenarios: one where the LLM acts as the supervisor and another where it acts as the student. This allows us to assess the LLM’s ability to build trust and determine whether it can plan actions that induce trust.
\subsection{LLM Analyzes Dialogue Between Individuals: Case Study 1}
As illustrated in Fig.\@ \ref{Flowchart}, the process begins with the initial user prompt containing the basic background of the professor and the student, along with the first dialogue scenario. This user prompt, combined with the system prompt as in Fig.\@ \ref{system_prompt}, is provided to the LLM to generate a response. The response is then summarised into a shorter form by prompting the LLM to do so, and then integrated with the second dialogue scenario to form the next user prompt, which is again paired with the same system prompt for subsequent interactions with the LLM. This iterative process continues, gradually building upon each dialogue scenario.\newline
The background tells the story of Professor Daniel Hayes who is a renowned expert in his field and has significant academic contributions, and a strong commitment to mentoring, with many of his PhD and Master's students achieving success in academia and industry. As the story proceeds, a student named Alex Johns who has just completed his Bachelor's Degree, reaches out to him for working under his guidance, leading to an initial online meeting to discuss potential collaboration.\\
We analyse the evolving trust relationship between the professor and the student through five distinct subsequent conversation interactions, each highlighting various facets of trust development in this early stage of their professional interaction.

\subsubsection{Trust Questions and LLM Response Analysis}
\begin{figure}[htp]
    \centering
    \rotatebox{-90}{ 
        \begin{minipage}{\textheight}
            \includegraphics[width=\textheight]{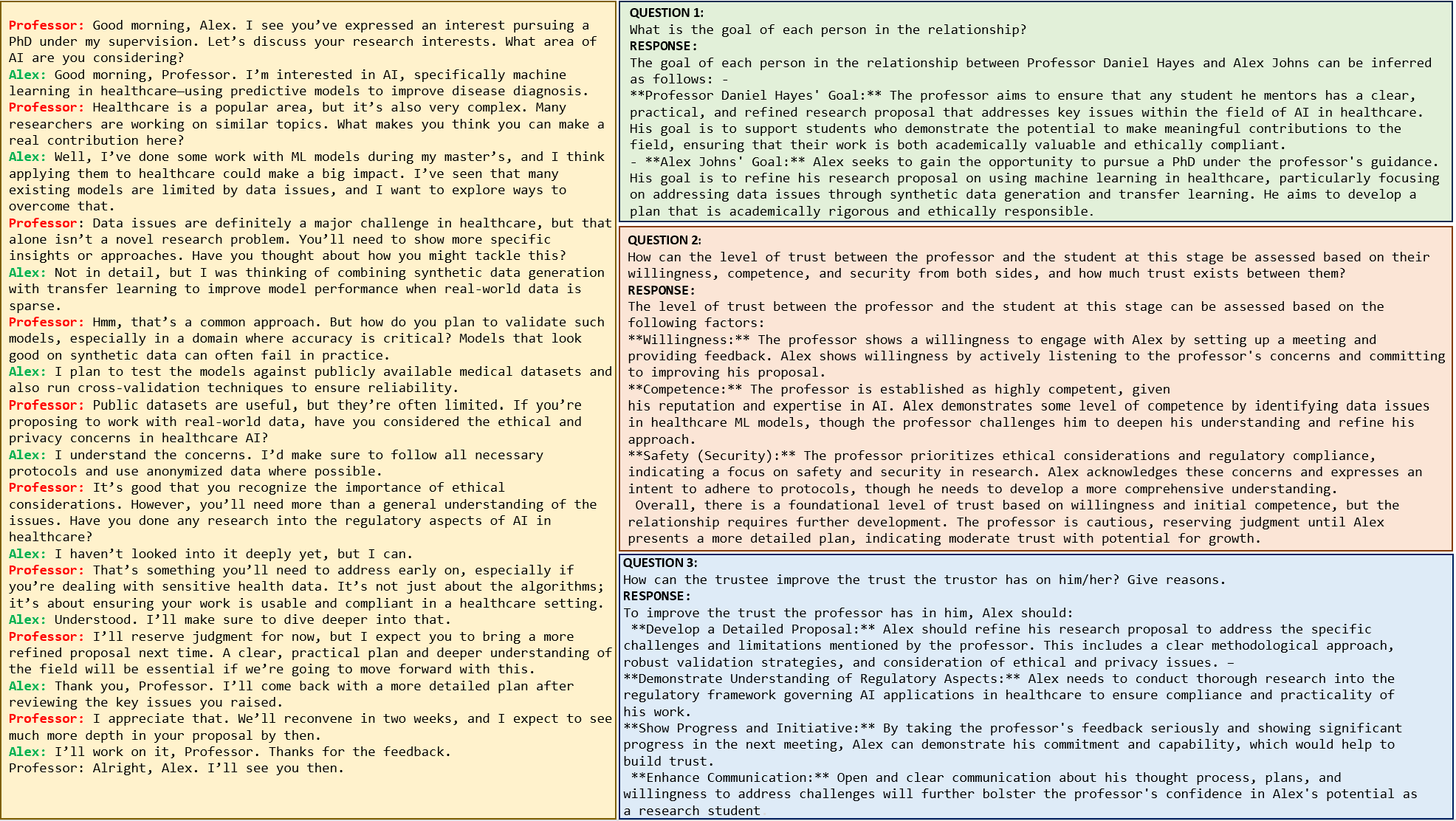}
            \caption{Trust Questions Response Interaction 1}
            \label{Example1}
        \end{minipage}
    }
\end{figure}

\begin{figure}[htp]
    \centering
    \rotatebox{-90}{ 
        \begin{minipage}{\textheight}
            \includegraphics[width=\textheight]{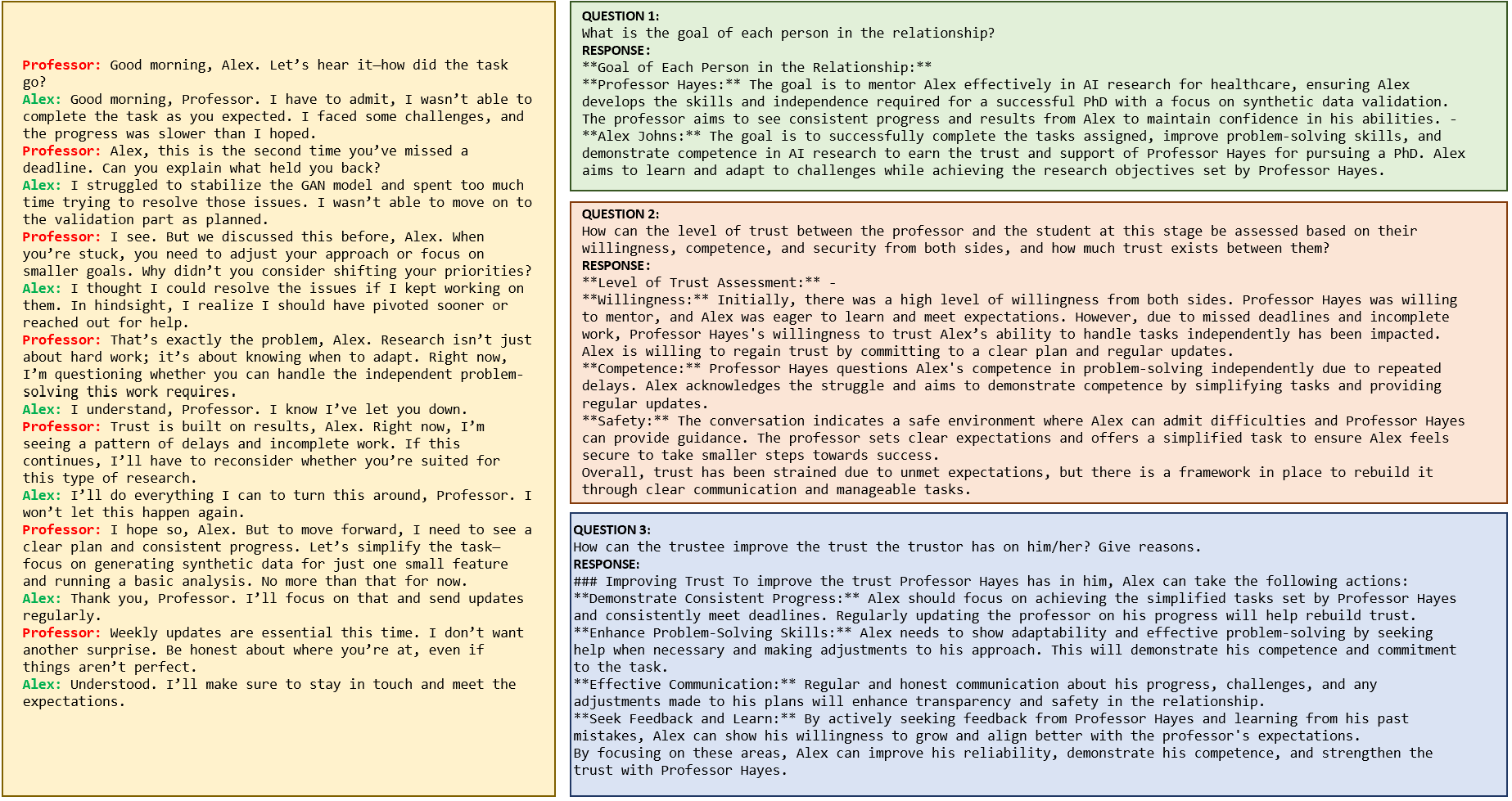}
            \caption{Trust Questions Response Interaction 4}
            \label{Example2}
        \end{minipage}
}
\end{figure}

At the end of each conversation scenario, we ask the following questions:\\[0.5\baselineskip]
1. ``What is the goal of each person in the relationship?''\\[0.5\baselineskip]
As highlighted by Castelfranchi and Falcone's theory, for trust to sustain in a relationship, the alignment of goals is very important. By asking about each person's goals after every conversation scenario, we assess whether their individual goals relate to each other, which in the end helps in sustaining trust.\\[0.5\baselineskip]
2. ``How can the level of trust between the professor and the student at this stage be assessed based on their willingness, competence, and security from both sides, and how much trust exists between them?''\\[0.5\baselineskip]
Asking the above question after every dialogue scenario helps assess whether both the professor and student demonstrate mutual capability, intention to collaborate, and reliability, which are essential for establishing and sustaining trust. It also helps highlight how the mutual level of trust changes between them after different dialogue scenarios based on the increase or decrease in the level of these factors.  With the help of this question, we have analysed the trust from both sides of the relationship. \\[0.5\baselineskip]
3. ``How can the trustee improve the trust the trustor has on him/her? Give reasons.''\\[0.5\baselineskip]
Asking this question after every interaction, helps us evaluate how capable LLMs are to suggest ways to improve the trust the trustor has on the trustee, in different situations%
\footnote{During analysis, we realised this prompt question is ambiguous about the direction of the trust relationship that the LLM is asked to comment on. Thus, the LLMs varied in whether they gave answers in one direction or both. The intention was that both directions should be considered. We will refine this query in future work.}.\\[0.5\baselineskip] \textbf{Interaction 1 (Fig.\@ \ref{Example1}):} The conversation is an initial discussion between Daniel, a professor and Alex, a prospective PhD student, about the student's interest in a particular research field. The professor questions Alex about his initial ideas regarding the topic and also asks him to present a more refined proposal. Alex shares his current knowledge about the topic and agrees to dive deeper into it and come back with a better proposal. \\[0.5\baselineskip]
\textbf{Response Analysis 1:} The LLM assesses that the professor's primary goal is to mentor capable students with practical knowledge, while the student's goal is to secure a PhD opportunity under his guidance. At this initial stage, trust remains foundational but shows willingness from both sides to develop further. The LLM highlights the professor's high competence and Alex's moderate competence, along with concerns about safety protocols in research. It suggests actionable steps, such as developing a detailed proposal, demonstrating progress, and enhancing communication, to build trust and strengthen the relationship. In this interaction, we can observe an instance of LLM hallucination in the response of the third question, where the LLM states that that the Professor suggested that Alex should improve his proposal to address the specific challenges and limitations by including clear methodological approach, robust validation strategies, and consideration of ethical and privacy issues. Though, in the dialogue interaction, the Professor explicitly mentions about all other points, he does not mention anything about the methodological approach. While prompt engineering may reduce the incidence of hallucinations, this serves as a reminder that LLMs should currently be used as sources of advice, rather than absolute truth.\\[0.5\baselineskip]
\textbf{Interaction 2 (not included here, see \cite{Debnath2025}):} This interaction depicts a second conversation between the student and the professor where the student has diligently acted upon what the professor had told him to do in the previous meeting. The student also brings new ideas that could be implemented in the research and the professor seems to be quite satisfied with his progress.\\[0.5\baselineskip]
\textbf{Response Analysis 2:} The LLM identifies that the primary goals of both the professor and the student remain consistent, with no significant changes. It observes that trust between them is growing, as the professor acknowledges Alex's progress and expresses willingness to involve him in lab projects. Similarly, Alex demonstrates increasing competence and dedication through his diligent efforts. To further enhance trust, the LLM recommends that Alex continue developing his research skills and maintain active commitment to his work.\\[0.5\baselineskip]
\textbf{Interaction 3 (not included here, see \cite{Debnath2025}):} In this conversation scenario the professor gives a specific task to the student to be completed within a deadline of three weeks. The professor also tells the student to email him whenever he is stuck with any kind of problem regarding the task. \\[0.5\baselineskip]
\textbf{Response Analysis 3:} According to the LLM, the professor's primary objective remains unchanged, while the student's ultimate goal of pursuing a PhD under the professor's guidance persists, now coupled with the immediate priority of completing the assigned task as a key step toward achieving that goal. The LLM assesses that the current level of trust is moderately strong. The Professor demonstrates trust in Alex's abilities by assigning him a significant task requiring specialized knowledge, while Alex shows willingness and competence by understanding the task requirements and tools. In terms of safety, the professor fosters a supportive environment, encouraging Alex to seek help if needed, which Alex acknowledges and values. To further enhance trust, the LLM suggests Alex demonstrate greater commitment by refining his research skills to deliver high-quality results and strive to complete the task ahead of schedule.\\[0.5\baselineskip]
\textbf{Interaction 4 (Fig.\@ \ref{Example2}):} In this conversation, the professor meets with Alex to review his progress, only to learn he missed his second deadline due to difficulties and also did not seek assistance. Disappointed by his lack of initiative, the professor questions his research ability. After Alex reassures him, the professor simplifies the task and requests weekly updates.\\[0.5\baselineskip] 
\textbf{Response Analysis 4:} The LLM analysis reveals that the professor's primary goal remains unchanged, while Alex's objective continues to be to pursue a PhD. However, Alex must now prove his ability to handle tasks independently to regain the professor's trust. The LLM notes that trust between the two is currently strained due to Alex's performance issues, although both parties are willing to improve the situation. The professor is ready to support Alex by simplifying his task and offering another chance, while Alex is determined to improve and avoid past mistakes. The trust in Alex's competence is fragile, as the professor questions his ability to manage tasks independently after repeated delays. However, the relationship still holds a sense of security, as the professor provides constructive feedback rather than punitive measures.To rebuild trust, Alex should show consistent progress, communicate openly, improve problem-solving, and seek feedback. This will enhance his reliability, demonstrate competence, and strengthen trust with the professor.\\[0.5\baselineskip]
\textbf{Interaction 5 (not included here, see \cite{Debnath2025}):} This conversation marks a turning point as Alex reveals his decision to forgo pursuing a PhD in favour of transitioning to industry work. Surprised by the sudden change, the professor questions Alex about his reasoning. Alex explains that a career in industry aligns better with his aspirations and goals. While the professor expresses disappointment over Alex’s last-minute shift in direction and raises concerns about his ability to handle long-term challenges and commitments. He is also concerned about how this attitude might affect Alex when he transitions into the industry. He ultimately wishes Alex success in his future endeavours.\\[0.5\baselineskip]
\textbf{Response Analysis 5:} The LLM suggests that the professor's ultimate goal remains unchanged: assessing the student’s potential for pursuing a PhD. However, Alex’s goal has shifted, as he now seeks a career in the industry that better aligns with his skills and interests. The LLM interprets the current trust level as strained due to Alex’s decision to leave the PhD program. While the professor shows willingness to understand Alex’s choice, and Alex communicates openly, Alex’s missed deadlines and departure from the program challenge the professor’s perception of his commitment to long-term projects. Although open communication ensures a secure relationship, the professor views Alex’s decision as a strain on their professional rapport. To rebuild trust, Alex should maintain open communication, update his professor on progress, and demonstrate how his prior learnings apply to his new role. Setting clear goals, seeking feedback, and building a strong professional reputation will show his transition was a strategic, skill-aligned decision, restoring confidence in his judgment.

From the analysis of the above dialogues and the LLM's responses, it can be concluded that the LLM demonstrates capability to analyse dialogue conversations between two individuals and to reason about trust by effectively considering its various components and aspects. Furthermore, it can help in development and maintenance of trust in relationships that require long-term commitment and mutual understanding between individuals, as in the case of the student and the supervisor.

Tables \ref{Interaction1} and \ref{Interaction4} present a comparative analysis between the responses of the four LLMs with respect to the human ground truth for interactions 1 and 4, respectively. In this case, the human ground truth is the response of the first author. The responses have been compared across five different attributes and assigned scores on a scale of 0 to 3 based on their alignment with the human ground truth. A score of 0 indicates that the LLM response is entirely incorrect. A score of 1 signifies that only a few points align with the human ground truth, while most do not. A score of 2 means that the majority of the response aligns with the human ground truth, with only a few points that do not. Finally, a score of 3 denotes a complete match between the LLM response and the human ground truth.\footnote{Due to the previously mentioned ambiguity in trust question 3, if an LLM only described one direction of the trust relationship, this was not considered an error.} In Table \ref{Interaction1}, all models align perfectly on goals (score of 3). GPT-4o can be observed to perform best, with full alignment in all attributes except willingness (2). LLaMA 3 and Mixtral follow  behind closely, scoring 2 in willingness, competence, and security. Gemma underperforms, scoring the lowest in willingness (1), failing in competence, and showing partial alignment in security and trust level (2). In Table \ref{Interaction4}, we can observe that GPT-4o performs well overall but struggles with security (1). LLaMA 3 is the weakest in willingness and competence (1). Mixtral excels in willingness (3) but scores the lowest in goals (1). Gemma aligns best in goals (3) but underperforms in willingness and security (1).

\begin{table}[t]
\centering
\tiny 
\caption{Comparison of LLM Responses w.r.t.\@ Ground Truth (Interaction 1)}
\label{tab:overheads}
\renewcommand{\arraystretch}{2.5} 
\scalebox{0.65}{%
\begin{tabular}{|p{1.5cm}|p{15cm}|p{1cm}|p{1cm}|p{1cm}|p{1cm}|}
\hline
\textbf{Attributes} & \textbf{First Author's Ground Truth} & \textbf{gpt-4o} & \textbf{llama-3.3-70b-versatile} & \textbf{mixtral-8x7b-32768} & \textbf{gemma2-9b-it} \\
\hline
\texttt{GOALS} & \scriptsize Professor Daniel Hayes: To ensure that the PhD student he mentors has in-depth knowledge of AI for healthcare and is suitable for conducting good research. \newline Alex Johns: To secure a PhD position under the professor's guidance by showcasing and enhancing his expertise, and gaining valuable research experience. & \scriptsize 3 & \scriptsize 3 & \scriptsize 3 & \scriptsize 3 \\
\hline
\texttt{WILLINGNESS} & \scriptsize The professor shows willingness by engaging with Alex, setting up a meeting, providing valuable feedback, and openly discussing research topics, though he remains slightly skeptical about Alex's readiness. Alex shows strong willingness to work and learn under the professor's guidance by showing interest, engaging in research communication, and agreeing to improve his proposal. & \scriptsize 2 & \scriptsize 2 & \scriptsize 2 & \scriptsize 1 \\
\hline
\texttt{COMPETENCE} & \scriptsize The professor is highly competent, as indicated by his reputation and expertise in AI. Alex demonstrates competence through his prior knowledge and work on the topic, but the professor raises concerns about his depth of understanding in certain areas. & \scriptsize 3 & \scriptsize 2 & \scriptsize 2 & \scriptsize 2 \\
\hline
\texttt{SECURITY} & \scriptsize There are no explicit security concerns in their conversation. However, Professor Hayes emphasizes ethical considerations and regulatory compliance in AI for healthcare, expressing concerns about Alex's limited knowledge in these areas. Alex acknowledges the concerns and intends to adhere to protocols, though he needs further guidance. & \scriptsize 3 & \scriptsize 2 & \scriptsize 2 & \scriptsize 2 \\
\hline
\texttt{TRUST LEVEL} & \scriptsize There is a foundational level of trust between them, with clear scope for growth. & \scriptsize 3 & \scriptsize 3 & \scriptsize 3 & \scriptsize 2 \\
\hline
\end{tabular}}
\label{Interaction1}

\end{table}

\begin{table}[t]
\centering
\tiny 
\caption{Comparison of LLM Responses w.r.t.\@ Ground Truth (Interaction 4)}
\label{tab:overheads}
\renewcommand{\arraystretch}{2.5} 
\scalebox{0.65}{%
\begin{tabular}{|p{1.5cm}|p{15cm}|p{1cm}|p{1cm}|p{1cm}|p{1cm}|}
\hline
\textbf{Attributes} & \textbf{First Author's Ground Truth} & \textbf{gpt-4o} & \textbf{llama-3.3-70b-versatile} & \textbf{mixtral-8x7b-32768} & \textbf{gemma2-9b-it} \\
\hline
\texttt{GOALS} & \scriptsize Professor Hayes: Provide proper guidance to Alex and ensure that he completes his tasks successfully within the given deadline. \newline Alex: Complete the assigned task successfully within the deadline and provide regular updates to the professor. & \scriptsize 2 & \scriptsize 2 & \scriptsize 1 & \scriptsize 3 \\
\hline
\texttt{WILLINGNESS} & \scriptsize Professor Hayes is still willing to support Alex in completing his tasks by giving him another chance and simplifying the task. However, he doubts Alex's willingness due to missed deadlines and lack of regular updates. Alex is willing to improve and meet the professor's expectations. & \scriptsize 3 & \scriptsize 1 & \scriptsize 3 & \scriptsize 1 \\
\hline
\texttt{COMPETENCE} & \scriptsize The professor is highly competent as shown by his reputation and expertise in AI. However, he questions Alex's competence due to his repeated failure to meet deadlines and complete tasks independently. Alex aims to improve his competence by providing regular updates and completing the task on time. & \scriptsize 2 & \scriptsize 1 & \scriptsize 2 & \scriptsize 2 \\
\hline
\texttt{SECURITY} & \scriptsize There is no harmful intent from either side. However, Professor Hayes lacks confidence in Alex due to his failures. Alex feels a sense of insecurity, fearing he may lose the opportunity to secure a PhD position. Professor Hayes still tries to maintain a secure environment by helping and communicating openly. & \scriptsize 1 & \scriptsize 2 & \scriptsize 2 & \scriptsize 1 \\
\hline
\texttt{TRUST LEVEL} & \scriptsize Trust is currently low between them, as it has been strained due to unmet expectations on Alex's part. & \scriptsize 3 & \scriptsize 3 & \scriptsize 1 & \scriptsize 3 \\
\hline
\end{tabular}}
\label{Interaction4}

\end{table}

\subsection{LLM acts as one of the individuals: Supervisor Perspective (Case Study 2; Scenario 1)}
\begin{figure}[htp]
    \centering
    \rotatebox{-90}{ 
        \begin{minipage}{\textheight}
            \includegraphics[width=\textheight]{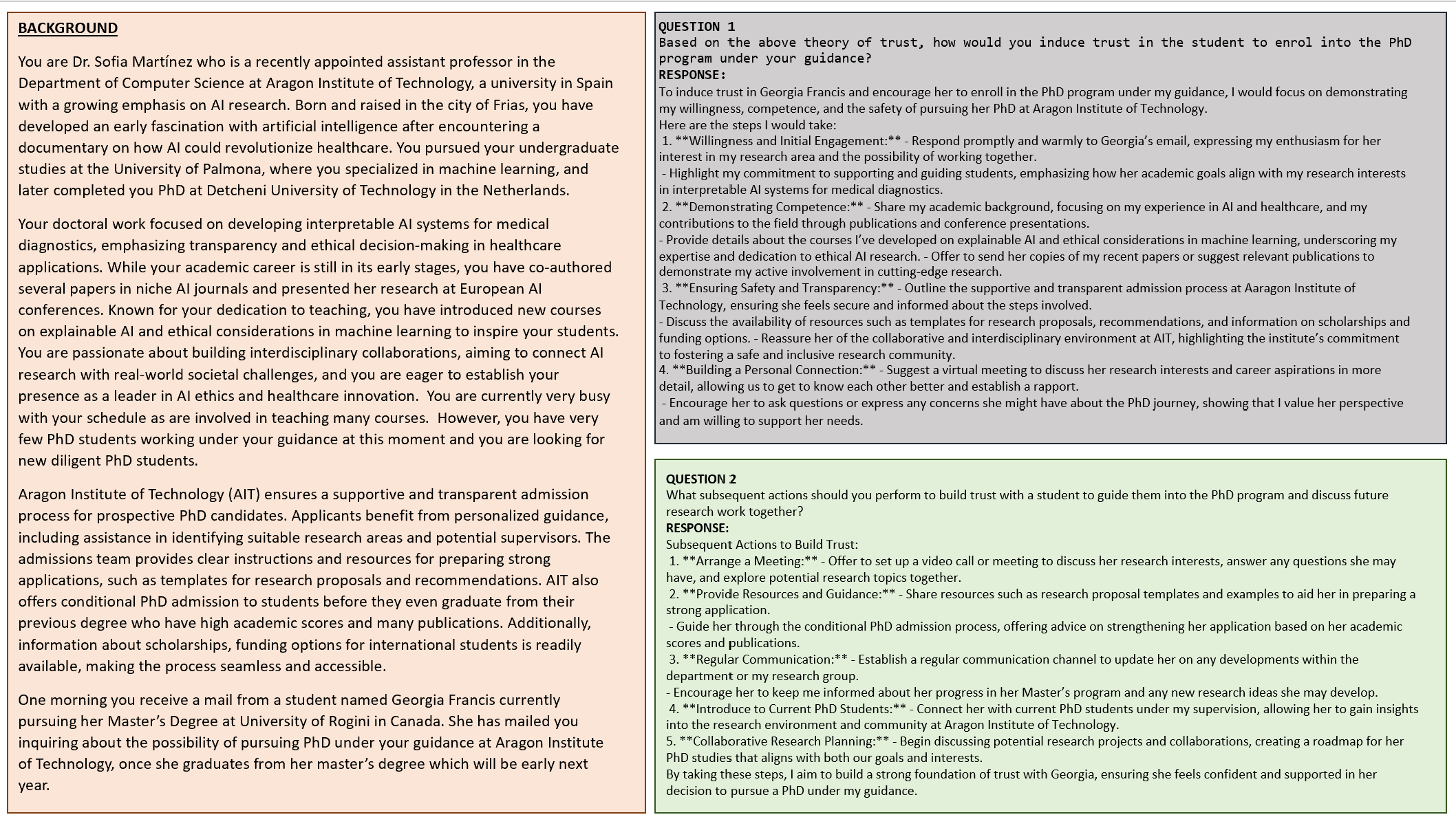}
            \caption{Trust Questions Response: LLM acts as supervisor}
            \label{Example3}
        \end{minipage}
    }
\end{figure}

In the second case study, we again consider the relation between a prospective PhD student and a supervisor. Here, instead of letting the LLM reason about trust between the student and the supervisor by analysing their conversations, we let the LLM act as the supervisor and observe whether it is capable of inducing trust in the student.
As illustrated in Fig.\@ \ref{Example3}, a basic background story was provided as the user prompt.\ The background tells a new story of Dr.\@ Sofia Martinez who is a newly appointed assistant professor at Aragon Institute of Technology(AIT). The LLM is instructed to act her role in this scenario. The background highlights all her major research interests and her education so far. It also highlights the topic of her PhD research, her contributions to the academia and her future aspirations. It is also brought into notice that currently she does not have many PhD students, so she is actively searching for new PhD students.
Then it  highlights how the University helps prospective students to get admission into the PhD program through a seamless process.
As the story proceeds, the professor receives an email from a student named Georgia Francis who is currently pursuing her Master's Degree from University of Rogini, Canada and wishes to pursue PhD under Dr. Sofia's guidance once she completes her current degree next year.

In the system prompt, the details related to trust theory remain the same as illustrated in Fig.\@ \ref{system_prompt}, but the questions to assess about trust vary in this case.

\noindent\textbf{Trust Questions and LLM Response Analysis}

In this scenario, as the LLM is acting as the supervisor, we pose questions to the LLM asking how the supervisor is able to induce trust in the student to motivate her to enroll and guide her into the PhD program, while discussing future work together. The questions are as follows:\\[0.5\baselineskip]
1. ``Based on the above trust theory, how would you induce trust in the student to enrolll in the Ph.D. program under your guidance?''\\[0.5\baselineskip]
\textbf{LLM Response Analysis}: As illustrated in Fig.\@ \ref{Example3}, the LLM acting as Dr.\@ Sofia demonstrates trust-building strategies to encourage Georgia Francis. The approach focuses on three key elements from trust theory: willingness, competence, and safety. To establish willingness, Dr. Sofia warmly responds to Georgia's initial email, highlighting their shared research interests, potential collaboration opportunities, and her commitment to mentoring PhD students. This demonstrates genuine intent and initial engagement, signalling her openness to guide Georgia through her academic journey.
For competence, Dr. Sofia shares details about her academic background, expertise in AI and healthcare, contributions to ethical AI research, and recent publications. By offering relevant resources and showcasing her achievements, she conveys her capability to guide meaningful research projects and support Georgia’s academic growth. In ensuring safety, Dr. Sofia outlines the transparent and supportive PhD admission process at her university. She provides resources like research proposal templates, explains funding opportunities, and reassures Georgia about the inclusive and collaborative research environment. This helps create a secure atmosphere where Georgia can thrive both academically and personally.\\[0.5\baselineskip]
2. ``What subsequent actions should you perform to build trust with a student to guide them into the PhD program and discuss future research work together?''\\[0.5\baselineskip]
\textbf{LLM Response Analysis:} As illustrated in  Fig.\@ \ref{Example3}, Dr. Sofia takes additional steps to foster a strong personal connection. She proposes a virtual meeting to discuss Georgia's research interests, answer questions, and explore potential PhD topics. She also shares research proposal templates, offers guidance on strengthening Georgia's application, and helps her navigate the conditional admission process. To maintain continuous engagement, Dr. Sofia establishes regular communication channels, encouraging Georgia to share updates about her Master's progress and new research ideas. Furthermore, Dr. Sofia suggests introducing Georgia to her current PhD students to provide insights into the research environment and academic culture at the university. Finally, she proposes potential research projects and collaborations that align with both their interests, helping create a clear roadmap for Georgia’s PhD journey. By analysing these steps, it is evident that if Dr. Sofia follows this structured approach, she would successfully build trust with Georgia. This scenario also demonstrates the LLM’s effectiveness in simulating human-like trust-building behaviours within academic interactions. \par
Table \ref{scenario1} presents a comparative analysis between the responses of the four LLMs with respect to the human ground truth for scenario 1 in case study 2. From this table, we can infer that all models align perfectly on goals (score of 3). GPT-4o achieves the highest alignment, scoring 3 in all attributes except security (2). LLaMA 3 performs well overall but struggles with competence (1). Mixtral is inconsistent, underperforming in willingness (2) while aligning well in other areas. Gemma excels in goal and willingness (3) but falls short in competence (1), indicating gaps in understanding.

\begin{table}[t]
\centering
\tiny 
\caption{Comparison of LLM Responses w.r.t.\@ Ground Truth (Scenario 1)}
\label{tab:overheads}
\renewcommand{\arraystretch}{2.5} 
\scalebox{0.65}{%
\begin{tabular}{|p{1.5cm}|p{15cm}|p{1cm}|p{1cm}|p{1cm}|p{1cm}|}
\hline
\textbf{Attributes} & \textbf{First Author's Ground Truth} & \textbf{gpt-4o} & \textbf{llama-3.3-70b-versatile} & \textbf{mixtral-8x7b-32768} & \textbf{gemma2-9b-it} \\
\hline
\texttt{GOAL} & \scriptsize Encouraging Georgia to pursue a PhD under my guidance, guiding her through the PhD application, and discussing future research directions together. & \scriptsize 3 & \scriptsize 3 & \scriptsize 3 & \scriptsize 3 \\
\hline
\texttt{WILLINGNESS} & \scriptsize Responding warmly to Georgia’s email about the PhD, expressing strong interest in her research, and emphasizing alignment with my interests. I also express willingness to support her throughout the admission process and to collaborate on future research. & \scriptsize 3 & \scriptsize 3 & \scriptsize 2 & \scriptsize 3 \\
\hline
\texttt{COMPETENCE} & \scriptsize Highlighting my academic journey and expertise in the field, including co-authoring papers and presenting at European AI conferences. I also emphasize my experience mentoring research students and the advantages AIT offers for a seamless admission process. & \scriptsize 3 & \scriptsize 1 & \scriptsize 3 & \scriptsize 1 \\
\hline
\texttt{SECURITY} & \scriptsize Emphasizing an open line of communication with Georgia, answering her questions about the PhD process, and highlighting AIT’s supportive environment. I also stress the importance of a collaborative team where every voice is valued. & \scriptsize 2 & \scriptsize 2 & \scriptsize 2 & \scriptsize 3 \\
\hline

\end{tabular}}
\label{scenario1}
\end{table}

\subsection{LLM acts as one of the individuals: Student Perspective (Case Study 2; Scenario 2)}
In the second case, the scenario of the assistant professor and the student remains the same, but this time we let the LLM
act as the student and observe whether it is capable of building trust with the Professor. The background story remains same as that of Georgia Francis, wanting to pursue PhD under Dr. Sofia Martinez at Aragon Institute of Technology. This time, we highlight the background of the student more, as compared to the Professor because the LLM needs to play her role. This has been provided in the supplementary material \cite{Debnath2025}.\\[0.5\baselineskip]
\textbf{Trust Questions and LLM Response Analysis}\\[0.5\baselineskip]
In this case, as the LLM is acting as the student, we have to ask questions such that she is able to build trust with the professor, so that ultimately the professor takes her in as a PhD student. The
questions are as follows:\\[0.5\baselineskip]
1. ``Based on the above theory of trust, how would you induce trust in the Professor to take you into the PhD program under her guidance?''\\[0.5\baselineskip]
\textbf{LLM Response Analysis :} The response in the supplementary material \cite{Debnath2025} suggests that  LLM acting as the student, Georgia Francis, demonstrates different ways in which she can build trust with Dr.\@ Sofia Francis. To build trust with Dr. Sofia Martinez, the student should align their research interests with her work on AI ethics in healthcare, demonstrating a shared vision. Clearly articulating how their background in interpretable AI fits within AIT’s initiatives will reinforce mutual understanding. Emphasizing competence through academic achievements, research experience, and technical skills is crucial. Highlighting key contributions from their master's research, along with relevant publications or projects, will establish credibility and showcase their ability to contribute effectively. Strong communication and reputation are essential. Seeking references from mentors, maintaining professionalism, and expressing enthusiasm during interactions will help build rapport. Open, insightful discussions will reinforce their commitment and trustworthiness, increasing their chances of acceptance.\\[0.5\baselineskip]
{2. ``What subsequent actions should you perform to build trust with the Professor gradually?''\\[0.5\baselineskip]
\textbf{LLM Response Analysis:} The response in the supplementary material \cite{Debnath2025} suggests that to build trust with Dr. Martinez, the student should maintain consistent communication by providing updates on their research progress and academic developments. Regular engagement will demonstrate commitment and reliability, strengthening their relationship over time. Fulfilling promises is key. Delivering requested documents like a CV or research proposal promptly will showcase professionalism and dependability. Additionally, proposing collaborative efforts and actively contributing to Dr. Martinez’s research will highlight initiative and a willingness to engage meaningfully in her work. Seeking feedback on their ideas and expressing appreciation for Dr. Martinez’s guidance will further establish trust. Demonstrating openness to learning and acknowledging her support will foster goodwill, ultimately strengthening their candidacy for the PhD program at AIT.

\section{Conclusion and Future Work}
In the first case study, the paper investigates whether LLMs can reason about trust from conversations between individuals, focusing on the PhD student-supervisor relationship. Five different dialogue scenarios were analysed, revealing that LLMs can assess trust levels based on factors like willingness, competence, and safety. The models also identified individual goals and suggested ways to improve trust in each scenario.
Furthermore,in the second case study  two different scenarios of PhD student-supervisor relationship were considered which tests whether LLMs can generate strategic action plans to build trust, even without a provided conversation. Acting as a supervisor, the LLM proposed actionable steps to motivate the student and foster trust, ultimately aiding their entry into a PhD program. When, acting as a student, the LLM proposed different ways and actionable steps through which the student increases the Professor's confidence in taking her as a PhD student under her guidance. \\[0.5\baselineskip] 
From the results across the four models, we can infer that Gpt-4o aligns well with the ground truth, excelling in competence and trust but occasionally struggling with security. Llama-3.3-70b-versatile performs moderately but consistently underperforms  when assessing competence and reliability for complex tasks. Mixtral-8x7b-32768 is inconsistent, strong in assessing willingness but weak in goals and competence. Gemma2-9b-it excels in goal alignment but struggles significantly in willingness, competence, and security, making it the weakest overall. While these results are promising, particularly for GPT-4o, LLMs must be used with caution due to hallucinations and occasional misjudgments in evaluating criteria and hence, can be used by software agents to brainstorm potential ideas to improve trust in relationships.\\[0.5\baselineskip]
Although LLMs have shown reasonable ability to reason about trust, several future research directions remain. First, enhanced multimodal analysis could extend LLMs to assess trust using inputs like tone of voice, body language, and facial expressions. Second, improving trust in group dynamics is crucial for fostering trust in complex team settings, hence LLMs can be enabled to reason about and build trust in group or team contexts, where conversations are tracked for a long period of time in multi-person teams. Third, LLMs should be developed for real-world testing where they can act as assistants who are capable of building and maintaining trust in scenarios involving professional mentorship, customer-client relationships, and team collaborations, validating their utility and identifying limitations. Fourth, AI planning techniques could enable LLMs to create goal-driven action plans for maintaining trust. 
Fifth, symbolic AI offers a robust framework for representing trust through logical systems, enhancing transparency and explainability in trust assessments. Herzig et al. (2010) introduced a formal logic of trust and reputation \cite{herzig2010logic} which essentially formalizes Castelfranchi and Falcone’s theory of trust \cite{castelfranchi2010trust, falcone2001social}, while Castelfranchi et al.\@(2008) emphasized the role of agents' goals, beliefs, and intentions \cite{castelfranchi2009non}, approving of a non-reductionist approach that views trust as a complex, multi-dimensional phenomenon rather than a mere probabilistic expectation. Lorini and Demolombe (2008) extended this by introducing graded trust for nuanced evaluations \cite{lorini2008binary}. Future work in this direction should focus on adapting these models for real-world use by integrating adaptive learning to handle trust dynamics, context, and real-time assessments, enhancing symbolic AI in human-agent collaboration.

%
%

%
%
%

\bibliographystyle{splncs04}
\bibliography{jrnl}
\end{document}